\documentstyle[aps,pra,twocolumn,epsfig]{revtex}

\draft

\begin{document}
\title{Direct measurement of the Wigner function by photon counting}
\author{K. Banaszek, C. Radzewicz, and K. W\'{o}dkiewicz}
\address{Wydzia{\l} Fizyki, Uniwersytet Warszawski,
Ho\.{z}a 69, PL-00-681 Warszawa, Poland}
\author{J. S. Krasi\'{n}ski}
\address{Center for Laser and Photonics Research, Oklahoma
State University, 413~NCR, Stillwater, OK~74078, USA}

\date{\today}

\maketitle

\begin{abstract}
We report a direct measurement of the Wigner function characterizing
the quantum state of a light mode. The experimental scheme is based on
the representation of the Wigner function as an expectation value of a
displaced photon number parity operator.  This allowed us to scan the
phase space point-by-point, and obtain the complete Wigner function
without using any numerical reconstruction algorithms.
\end{abstract}


Among many representations of the quantum state, the Wigner function
offers an appealing possibility to describe quantum phenomena using the
classical-like concept of phase space \cite{WignerFunction}.  The Wigner
function provides complete information on the state of a system, and it
allows one to evaluate any quantum observable by phase space integration
with an appropriate Wigner-Weyl ordered expression.

Recently, the Wigner function has gained experimental significance
due to the development of the optical homodyne tomography, a beautiful
technique for measuring the quantum state of light pioneered by Smithey
{\em et al.} \cite{SmitBeckPRL93} and further applied by Breitenbach
{\em et al.} \cite{BreiSchiNAT97} In this method, rooted in the domain
of image processing, the Wigner function is a natural representation
of the quantum state reconstructed from experimental data. However,
the route from raw experimental results to the Wigner function is not
straightforward. First, a sample of homodyne events is collected and
stored. Statistics of these events for a fixed local oscillator phase is
described by a marginal projection of the Wigner function. In order to
retrieve the complete Wigner function, a family of homodyne statistics
measured for a sufficiently dense set of local oscillator phases  has to
be processed using the sophisticated filtered back-projection algorithm.

In this Communication we report a direct measurement of the Wigner
function of a light mode. This technique, based on photon counting,
avoids the detour via complex numerical reconstruction algorithms.  The
principle of our measurement is entirely different from optical homodyne
tomography.  The Wigner function at a given phase space point is itself
a well defined quantum observable \cite{WignerandParity}. Furthermore,
the measurement of this observable can be implemented for optical
fields using an arrangement employing an auxiliary coherent probe
beam \cite{WallVogePRA96,BanaWodkPRL96}. The amplitude and the phase
of the probe field define the point in the phase space at which the
Wigner function is measured. This allowed us to scan the phase space
point-by-point, simply by changing the parameters of the probe field. A
variation of this idea has been applied by Leibfried {\em et al.}
\cite{LeibMeekPRL96} to determine the vibrational state of a trapped
ion. Here we present an experiment, which to the best of our knowledge
is the first direct measurement of the Wigner function for optical fields.

Our experiment is based on the representation of the Wigner function
at a complex phase space point denoted by $\alpha$ as the expectation
value of the following operator:
\begin{equation}
\label{Eq:Wdef}
\hat{W}(\alpha) = \frac{2}{\pi}
\sum_{n=0}^{\infty} (-1)^{n} \hat{D}(\alpha)|n\rangle \langle
n | \hat{D}^{\dagger}(\alpha),
\end{equation}
where $\hat{D}(\alpha)$ is the displacement operator and $|n\rangle$
denote Fock states, $\hat{n}|n\rangle = n|n\rangle$.
Thus, $\hat{W}(\alpha)$ has two eigenvalues:
$2/\pi$ and $-2/\pi$, corresponding to degenerate subspaces spanned
respectively by even and odd displaced Fock states.  Practical means
to translate this formula into an optical arrangement are quite simple
\cite{WallVogePRA96,BanaWodkPRL96}.  The displacement transformation can
be realized by superposing the measured field at a low-reflection beam
splitter with a strong coherent probe beam. The value of the displacement
$\alpha$ is equal in this setup to the reflected amplitude of the probe
field. Furthermore, the projections on Fock states can be obtained by photon
counting assuming unit quantum efficiency. These two procedures, combined
together, provide a practical way to measure the Wigner function at an
arbitrarily selected phase space point $\alpha$.

\begin{figure}
\noindent\centerline{\epsfig{file=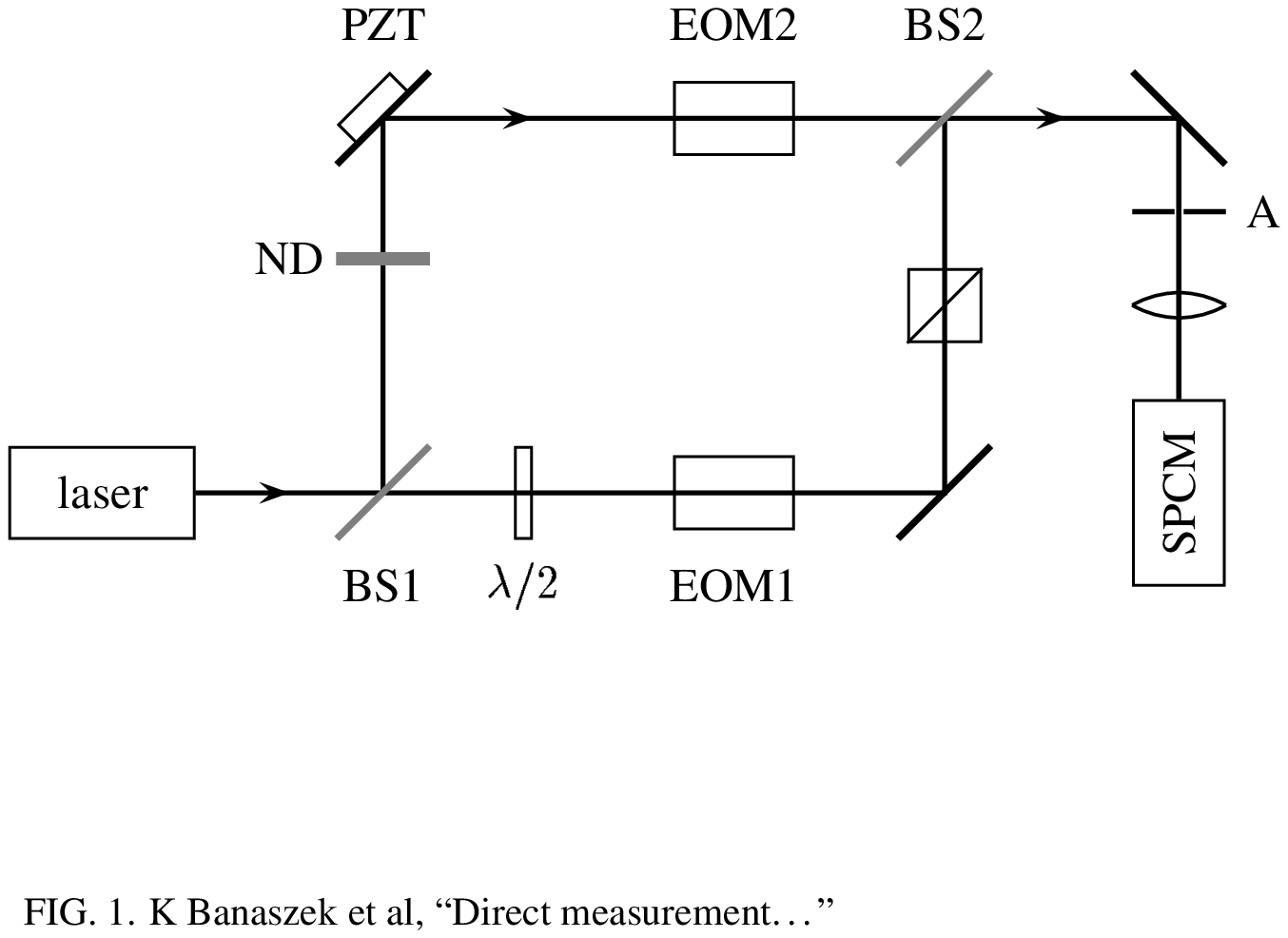,clip=,angle=90,width=3.375in}}

\bigskip

\caption{The experimental setup for measuring the Wigner function.
BS1 and BS2 are quartz plates serving as low-reflection beam splitters.
The quantum state is prepared using the neutral density filter ND and
a mirror mounted an a piezoelectric translator PZT. The
electrooptic modulators EOM1 and EOM2 control respectively the amplitude
and the phase of the point at which the Wigner function is measured. The
signal field, after removing spurious reflections using the aperture A,
is focused on a single photon counting module SPCM.}
\label{Fig:Setup}
\end{figure}

The experimental setup we used to measure the Wigner function is
shown schematically in Fig.~\ref{Fig:Setup}.  In principle, it is a
Mach-Zender interferometric scheme with the beams in two arms
of the interferometer serving as the
signal and the probe fields.  An attenuated, linearly polarized 
(in the plane of Fig.~\ref{Fig:Setup}) 632.8~nm
beam from a frequency-stabilized single-mode He:Ne laser is divided by
a low-reflection beam splitter BS1. The weak reflected beam is used
to generate the signal field whose Wigner function will be measured.
The state preparation stage consists of a neutral density filter ND
and a mirror mounted on a piezoelectric translator PZT. With this
arrangement, we are able to create pure coherent states with variable
phase as well as their incoherent mixtures. Though these states do not
exhibit nonclassical properties, they constitute a nontrivial family
to demonstrate the principle of the method, which provides a complete
characterization of both quantum and classical field fluctuations.

The strong beam leaving the beam splitter BS1 plays the role of the
probe field with which we perform the displacement transformation
$\hat{D}(\alpha)$. In order to scan the phase space one should be able
to set freely its amplitude and phase, which define respectively the
radial and the angular coordinates in the phase space. The amplitude
modulation is achieved with a half-wave plate, a longitudinal Pockels
cell EOM1, and a polarizer oriented parallel to the initial direction
of polarization. The phase modulation is done with the help of an ADP
crystal electrooptic phase modulator EOM2 on the signal field.  This is
completely equivalent to modulating the probe field phase, but more
convenient for technical reasons: in this arrangement optical paths in
both the arms of the Mach-Zender interferometer are approximately the
same, and better overlap of the signal and the probe modes is achieved
at the output of the interferometer.

The signal and the probe fields are interfered at a nearly
completely transmitting beam splitter BS2 with the power transmission
$T=98.6\%$. In this regime, the transmitted signal field effectively
undergoes the required displacement transformation.  
Spurious reflections
that accompany the beam leaving the interferometer are removed using
the aperture A. Finally, the transmitted signal is focused on an EG\&G
photon counting module SPCM-AQ-CD2749, whose photosensitive element is a
silicon avalanche diode operated in the Geiger regime. The overall
quantum efficiency of the module specified by the manufacturer is
$\eta \ge 70\%$.  The count rate is kept low in the experiment, and thus
the chance of two or more photons triggering a single avalanche signal
is very small, and the probability of another photon arriving during
the detector dead time can be neglected. Under these assumptions, each
pulse generated by the module corresponds to the detection of a single
photon \cite{DarkCounts}. The pulses are acquired by a computer, which
also controls the voltages applied to the electrooptic modulators. The
interference visibility in our setup has been measured to be $v \ge
98.5\%$, and the phase difference between the two arms was stable up to
few percent over times of the order of ten minutes.

\begin{figure}
\noindent\epsfig{file=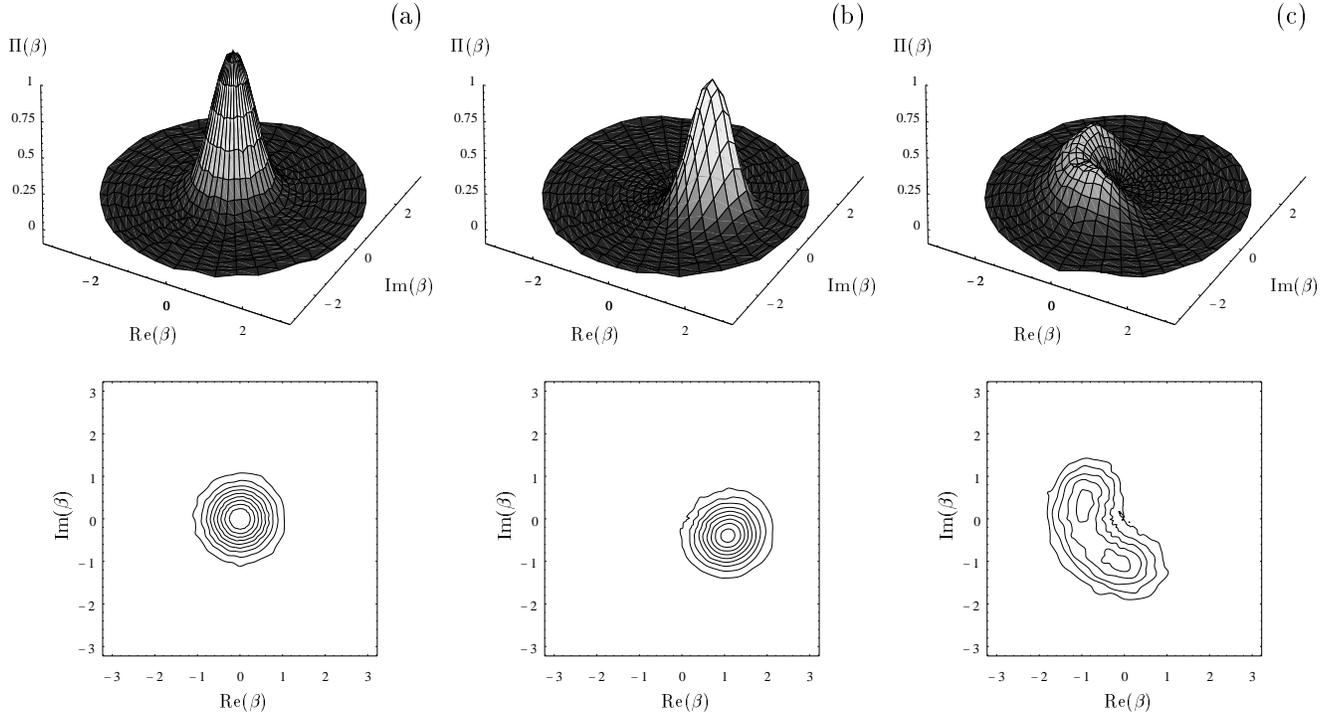,angle=90,width=7in,clip=}

\bigskip

\caption{The measured Wigner functions of (a) the vacuum, (b) a weak
coherent state, and (c) a phase diffused coherent state. The contour plots
depict interpolated heights given by multiples of 0.1 for the
plots (a) and (b), and by $0.08, 0.14, 0.20, 0.26, 0.32$ for the plot
(c).}
\label{Fig:Wigner}
\end{figure}

In Fig.~\ref{Fig:Wigner} we depict the measured Wigner functions of the
vacuum, a weak coherent state, and a phase diffused coherent
state.  The phase fluctuations were obtained by applying a 400~Hz
sine waveform to the piezoelectric translator. For all the plots,
the phase space was scanned on a grid defined by 20 amplitudes and
40 phases. The scaling of the radial coordinate is obtained from the
average number of photons $n_{\text{vac}}$ detected for the blocked
signal path. Thus the graphs are parameterized with the complex variable
$\beta=e^{i\varphi}n_{\text{vac}}^{1/2}$, where $\varphi$ is the
phase shift generated by the phase modulator EOM2.  At each selected
point of the phase space, the photocount statistics $p_n(\beta)$
was determined from a sequence of $N=8000$ counting intervals, each
$\tau=30\mu$s long. The duration of the counting interval $\tau$ defines
the temporal envelope of the measured mode. The count statistics was
used to evaluate the alternating sum
\begin{equation}
\Pi(\beta) = \sum_{n=0}^{\infty} (-1)^{n} p_n(\beta),
\end{equation}
which, up to the normalization factor $2/\pi$ is equal to the Wigner
function of the measured state.
Statistical variance of this result can be estimated
by $\text{Var}[\Pi(\beta)] = \{1-[\Pi(\beta)]^{2}\}/N$
\cite{BanaWodkJMO97}. Thus, the statistical error of our measurement
reaches its maximum value, equal to $1/N^{1/2} \approx 1.1\%$, when
the value of the Wigner function is close to zero.

The Wigner functions of the vacuum and of the coherent state are
Gaussians centered at the average complex amplitude of the field, and
their widths characterize quantum fluctuations. It can be noticed that
the measured Wigner function of the coherent state is slightly lower than
that of the vacuum state. In the following, when discussing experimental
imperfections, we shall explain this as a result of non-unit interference
visibility.  In the plot of the Wigner function of the phase diffused
coherent state, one can clearly distinguish two outer peaks corresponding
to the turning points of the harmonically modulated phase.

There are several experimental factors whose impact on the result of
the measurement needs to be analyzed. First, there are losses of the
signal field resulting from two main sources: the reflection from the
beam splitter BS2 and, what is more important, imperfect photodetection
characterized by the quantum efficiency $\eta$.  Analysis of these losses
shows \cite{WallVogePRA96,BanaWodkPRL96}, that in such a case the
alternating series evaluated from photocount statistics is proportional
to a generalized, $s$-ordered quasidistribution function $W(\alpha; s)$,
with the ordering parameter equal $s=-(1-\eta T)/\eta T$.

In addition, the two modes interfered at the beam splitter BS2 are never
matched perfectly. The effects of the mode mismatch can be discussed most
thoroughly within the multimode theory \cite{MultimodeTheory}. Here,
due to limited space, we shall present the main conclusions and briefly
sketch the reasoning. Let us consider the normalized mode functions
describing the transmitted signal field and the reflected probe field. The
squared overlap $\xi$ of these two mode functions can be related to the
interference visibility $v$ as $\xi=v/(2-v)$. In order to describe the effects
of the mode mismatch, we will decompose the probe mode function into a part
that precisely overlaps with the signal, and the orthogonal remainder.
The amplitude of the
probe field effectively interfering with the signal is thus multiplied
by $\xi^{1/2}$, and the remaining part of the probe field contributes to
independent Poissonian counts with the average number of detected photons
equal $(1-\xi)|\beta|^2$. Consequently, the full count statistics is
given by a convolution of the statistics generated by the interfering fields,
and the Poissonian statistics of mismatched photons. A simple calculation
shows, that the alternating sum evaluated from such a convolution
can be represented as a product of the contributions corresponding
to the two components of the probe field:
\begin{eqnarray}
\Pi (\beta)
& = & \exp[-2(1-\xi)|\beta|^2] \nonumber \\
& &
\times \frac{\pi}{ 2\eta T} 
W \left( \sqrt{\frac{\xi}{\eta T}} \beta ; - \frac{1-\eta T}{\eta T}
\right).
\end{eqnarray}
Here on the right-hand side we have made use of the
theoretical results for imperfect detection obtained in
Refs.~\cite{WallVogePRA96,BanaWodkPRL96}.
Specializing this result to
a coherent state $|\alpha_0\rangle$ with the amplitude $\alpha_0$,
yields:
\begin{equation}
\Pi(\beta) = \exp[-2|\beta - \sqrt{\xi \eta T } \alpha_0|^2
- 2 (1-\xi)\eta T |\alpha_0|^2].
\end{equation}
Thus, in a realistic case 
$\Pi(\beta)$ represents a Gaussian centered at the
attenuated amplitude $\sqrt{\xi \eta T }\alpha_0$, and the width
remains unchanged. This Gaussian function in multiplied by the constant
factor $\exp[- 2 (1-\xi)\eta T |\alpha_0|^2]$.
For our measurement, $\xi \approx 97\%$ and 
$\eta T |\alpha_0|^2 \approx 1.34$,
which gives the value of this factor equal 0.92. This result agrees
with the height of the
experimentally measured Wigner function of a coherent state.

We shall conclude this Communication with a brief comparison of the
demonstrated direct method for measuring the Wigner function with
the optical homodyne tomography approach. An important parameter
in the experimental quantum state reconstruction is the detection
efficiency. Currently, higher values of this parameter can be achieved
in the homodyne technique,
which detects quantum fluctuations as a difference between two
rather intense fields. Such fields can be efficiently converted into
photocurrent signals with the help of {\it p-i-n} diodes. It should
be also noted that an avalanche photodiode is not capable of resolving
the number of simultaneously absorbed photons, and that it delivers a
signal proportional to  the light intensity only in the regime described
in this paper. However, continuous progress in single photon detection
technology gives hope to overcome current limitations of photon counting
\cite{KwiaSteiPRA93}.  Alternatively, the displacement transformation
implemented in the photon counting technique can be combined with
efficient random phase homodyne detection. This yields the recently
proposed scheme for cascaded homodyning \cite{KisKissPRA99}.

The simplicity of the relation (\ref{Eq:Wdef}) linking the count
statistics with quasidistribution functions allows one to determine
the Wigner function at a given point from a relatively small sample of
experimental data.  This feature becomes particularly advantageous, when
we consider detection of multimode light.  Optical homodyne tomography
requires substantial numerical effort to reconstruct the multimode Wigner
function. In contrast, the photon counting method has a very elegant
generalization to the multimode case: after applying the displacement to
each of the involved modes, the Wigner function at the selected point
is simply given by the average parity of the total number of detected
photons. Moreover, the dichotomic outcome of such a measurement provides
a novel way of testing quantum nonlocality exhibited by correlated states
of optical radiation \cite{BanaWodkPRL99}.

The authors thank Prof.\ K. Ernst for placing a single-mode He:Ne laser
at their disposal. This research is supported by Komitet Bada\'{n}
Naukowych, Grant 2P03B~002~14.

\onecolumn

\begin{references}
\bibitem{DarkCounts}
The operating mode of the SPCM results in a certain amount of  extraneous
pulses originating from dark counts and afterpulsing. The dark count
rate of our module is less than 100~s$^{-1}$. This gives on average $<
3 \cdot 10^{-3}$ during a single counting interval, which is 30~$\mu$s
long. The typical afterpulsing probability is $0.2\%$.  Thus, both these
effects give a negligible contribution to the measured count statistics.
\end{references}
\end{document}